\documentclass{article}
\usepackage{amsmath}
\usepackage{amssymb}
\usepackage{amscd}

\def\sc{$\Sigma_c$}

\def\beq{\begin{equation}}
\def\eeq{\end{equation}}
\def\ba{\beq\begin{array}{l}}
\def\ea{\end{array}\eeq}
\def\be{\ba}
\def\ee{\ea}

\def\p{\partial}

\def\x{\alpha}
\def\y{\lambda}

\begin{document}
\begin{titlepage}
\today
\vskip 5 cm
\begin{center}
\large{\Large\bf{Black Holes as Incompressible Fluids on the Sphere}}
\vskip 0.5in
{Irene Bredberg 
and Andrew Strominger}
\vskip 0.3in
{\it Center for the Fundamental Laws of Nature,
Harvard University\
\
Cambridge, MA, 02138

}

\end{center}

\vskip 0.6cm

\begin{abstract}
We consider finite deformations of the $p+2$-dimensional Schwarzschild geometry which obey the vacuum Einstein equation,  preserve the mean curvature and induced conformal metric on a sphere a distance $\y$ from the horizon and are regular on the future horizon. We show perturbatively that in the limit $\y\to 0$ the deformations are given by solutions of the nonlinear incompressible Navier-Stokes equation on the $p$-sphere. This relation provides a link between global existence 
for $p$-dimensional incompressible Navier-Stokes fluids and a novel form of cosmic censorship in $p+2$-dimensional general relativity. 
\end{abstract}

\vspace{3.0cm}

\end{titlepage}

\setcounter{tocdepth}{2}
\tableofcontents

\section{Introduction}
    Inspired by earlier works spanning several decades \cite{thesDamour,oldDamour,Price:1986yy,Policastro:2001yc,minw}, a suitable definition of the intrinsic dynamics of planar Rindler-like horizons was recently found \cite{Bredberg:2010ky,bkls, Compere:2011dx,vlas} in which they are manifestly governed by the incompressible Navier-Stokes equation. The definition involved first putting a Dirichlet-like  boundary condition on the induced metric on a timelike hypersurface \sc\ just outside the horizon. This isolates the dynamics of the horizon from the bulk gravity waves and other excitations of the rest of the spacetime. When \sc\ is close enough, the interior excitations which remain after imposing this boundary condition can be viewed as intrinsic to the horizon. One then demands regularity on the future horizon and takes the mean curvature $K$ of  \sc\ to diverge in such a way that \sc\ approaches its Rindler horizon. The leading terms in the near-horizon expansion  in (roughly in $K^{-1}$) of the resulting smooth geometries are simply given by solutions of the incompressible Navier-Stokes equation on the plane.

   In this paper we generalize the construction of \cite{bkls} so that it is relevant to the spherical horizons of asymptotically flat Schwarzschild black holes in $p+2$ spacetime dimensions. After imposing the isolating  near-horizon boundary condition  and taking the near-horizon limit, we find that the dynamics of Schwarzschild horizons are governed by the incompressible Navier-Stokes equation on the sphere\footnote{We analyze the first three orders in perturbation theory but do not herein generalize the all-orders proof of \cite{Compere:2011dx}. Fluid flow on the sphere is also considered in Appendix B of \cite{minw}.}.

  The generalization of the planar analysis of \cite{bkls} to the spherical case herein was largely straightforward except for one surprise: at third order, a global obstruction to constructing the perturbed solution with fully Dirichlet boundary conditions is  encountered.   This obstruction (detailed below) is 
related to total energy conservation and is circumvented by allowing the conformal factor of the \sc\ metric to fluctuate while instead fixing the mean curvature $K$. Such a subleading modification of the boundary conditions was described in \cite{vlas} and does not affect the universal emergence of the incompressible Navier-Stokes dynamics in the near-horizon scaling limit.

   In our previous analysis of planar horizons \cite{bkls} a second limit, distinct but ultimately equivalent to the near-horizon limit,  was considered in which the data on \sc\ underwent a long-wavelength hydrodynamic scaling. There is no analog of this limit for fixed-mass Schwarzschild, because for any fixed sphere, going to wavelengths longer than the Schwarzschild radius leads to a trivial theory. Nevertheless it is still the case that the near-horizon limit discussed herein is well-defined and equivalent to a fluid in the usual hydrodynamic limit. 
   \section{Navier-Stokes on the $p$-sphere}
The form of the Navier-Stokes equation in flat space is highly constrained by Galilean invariance under the inertial transformation $x^i \to x^i +w^it, ~~t\to t$, where $w^i$ is a constant velocity. This symmetry in fact fixes the coefficient of the nonlinear term. 
In a general spatially curved background, there is no such symmetry and there are families of generalizations of the equations which 
differ in the coupling of $v^i$ to the curvature. However on the $p$-sphere with the round radius-$L$ metric $g_{ij}$ it is natural to demand that the equations are solved (with $P=v^2/2$) when $v^i$ generates a Killing symmetry of the sphere. This requirement leads to
\beq \label{ghy}
\p_t v^i + v^j \nabla_j v^i+ \nabla^i P - \nu \left(\nabla^2 v^i + R^i_j v^j\right)=0, ~~~ \nabla_i v^i=0,
\eeq
where $L^2 R_{ij}=(p-1)g_{ij}$ and 
$\nabla_kg_{ij}=0$ as usual. The viscosity $\nu$ is dimensionful and can be set to unity by rescaling $t,~v^i$ and $P$ while keeping the radius of the sphere fixed. Below we choose conventions in which  $\nu=1$.
\section{Schwarzschild}
The Schwarzschild metric in $p+2$ spacetime dimensions is 
\beq
ds_{Schw}^2 = -\left[1 - \left(\frac{2m}{r}\right)^{(p-1)} \right] d\tau^2 + \left[1 - \left(\frac{2m}{r}\right)^{(p-1)} \right]^{-1} dr^2 + r^2 d\Omega_p^2
\eeq
where $d\Omega_p^2$ is the round metric on the unit $p$-sphere. 
After transforming to ingoing null  coordinates  ($ t, \rho$) defined by
\beq
t= \frac{(p-1)\x  } {2 m }\left(\tau+  \large\int^r \frac{dr'}{ 1 - \left(\frac{2m}{r'}\right)^{(p-1)} }\right),~~~~\rho =\frac{2m}{\x (p-1)}(r-2m),
\eeq
the metric becomes
\begin{align}\label{nse}
ds_{Schw}^2 &= -\left(\frac{\rho}{\x  }-\frac{p(p-1 )\rho^2}{8 m^2} +  \frac{\x p(p+1) (p-1)^2  \rho^3}{96 m^4}-\frac{{\x}^2 p (p-1)^3(p+1)(p+2) \rho^4}{1536 m^6}\right) dt^2  \\
&\nonumber+2dtd\rho +4 m^2\left(1 + \frac{\x (p-1)  \rho}{2 m^2}+\frac{{\x}^2 (p-1)^2 \rho^2}{16 m^4}\right) d\Omega_p^2
 + O({\x}^3).
\end{align}
The advantage of these coordinates is that they are regular on the future horizon $\rho=0$. 
For $m\to \infty$ we recover a wedge of flat Minkowski space. We take our boundary hypersurface \sc\  to be located at $\rho=1$. 
\sc\ is pushed towards the horizon for $\x \rightarrow 0$.
\section{Deformed solution} 
Now we wish to perturbatively construct a deformed solution of the Einstein equation holding fixed the conformally round intrinsic geometry and mean curvature $K$ (the trace of the extrinsic curvature) of \sc. We also demand regularity on the future horizon,  $i.e.$ the metric coefficients are regular in the $(t,\rho)$ coordinates. We then consider the limiting behavior as $\x \to 0$, which implies that $K\to \infty$  and \sc\ approaches the horizon.  Fixing such Dirichlet boundary conditions on \sc\  shields the black hole from 
gravity waves and other perturbations of the ambient spacetime and isolates its internal dynamics. Deformations which survive this limit can be thought of as intrinsic to the black hole itself. 

We wish to do a large $K$ expansion of the full bulk geometry.  Our small expansion parameter $\y$ is defined to be proportional to the inverse square of the curvature
\beq
\y\equiv\frac{1}{4 K^2}=\frac{m^2 \left(1 + \frac{\x (p-1)}{4 m^2}\right)^2 \left[1 -\left(1 + \frac{\x (p-1)}{4 m^2}\right)^{-(p-1)}\right]}{\left[p - \frac{(p+1)}{2} \left(1 + \frac{\x (p-1)}{4 m^2}\right)^{-(p-1)}\right]^2}=\x + \mathcal{O}(\x^2).
\eeq
The first four terms in the $\y$ expansion of the smooth deformed solution obeying the specified boundary conditions are
\begin{align}
&ds^2 = -\frac{\rho}{\y}dt^2 \\&\nonumber+\frac{p (p-1)\rho \left(5+\rho\right)}{8 m^2}dt^2 +  2dtd\rho +4 m^2d\Omega_p^2
+\left(1 - {\rho} \right) \left[ v^2 dt^2 - 2 v_i dt dx^i \right]-2\rho P dt^2\\&\nonumber
+\y \Bigl[\left(2(p-1)\rho+ 8 m^2 P\right) d\Omega_p^2+(1-\rho)  {v_i v_j} dx^i dx^j\\&\nonumber-{\left(\rho^2-1\right)} \left(\nabla^2 v_i+ 
 (p-1) R_{j i} v^j \right) dtdx^i -2 v_i d\rho dx^i + \left(v^2 + 2 P\right){dt d\rho}  + 2    \left(1 - {\rho}\right) \phi_i dt dx^i\Bigr]  \\&\nonumber
 +\y^2 \Bigl(2 \chi_i dx^id\rho + \lambda_{ij} dx^idx^j\Bigr) +....
\end{align}
$\nabla_k, R_{ij}$ are the covariant derivative and Ricci tensor associated with the $p$-dimensional spatial $g_{ij}$ metric induced on $\Sigma_c$. Also, here $(v_i,P,\phi_i)$ depend only on $(t,x^i)$ and $(\chi_i,\lambda_{ij})$ are both functions of $(t,\rho,x^i)$ - these only appear in $G_{tt}$ at $\mathcal{O}(\y^0)$. The induced metric on a $\rho=1$ hypersurface is related via a conformal factor of 
\beq
1 + 2 \y P + \mathcal{O}(\y^2)
\eeq
to the direct product of the round metric on a radius $r_c$ $p$-sphere with a flat timelike line: 
\beq
-\left[\frac{1}{\y}+\frac{3 p (p-1)}{4 m^2}+\mathcal{O}(\y)\right]dt^2 + r_c^2d\Omega_p^2
\eeq
where $r_c\equiv \left.r\right|_{\rho=1}$ is $2m+ \mathcal{O}(\y) $. 

We find that for the perturbed geometry to satisfy the vacuum Einstein equations through $\mathcal{O}(\y^0)$, the pair $(v_i,P)$ need to solve the incompressible Navier-Stokes equation (\ref{ghy}) with viscosity $\nu=1$. The Navier-Stokes condition appears at $\mathcal{O}(\y^0)$ whilst incompressibility is necessary already at $\mathcal{O}(\y^{-1})$. Requiring that the radially independent contributions to $G_{tt}$ are solved through $\mathcal{O}(\y^0)$ fixes the divergence of $\phi_i(t,x^j)$ (we have used incompressibility and the Navier-Stokes equation to rewrite this condition)
\beq\label{phie}
\nabla_i\phi^i=p\p_tP + \p_t(v^2) +\nabla_i\left[(p-1)P v^i -\frac{1}{2}\nabla^2v^i-v^j\nabla_jv^i-v^j\nabla^iv_j\right].
\eeq
Solving for the entire $G_{tt}$ to vanish through $\mathcal{O}(\y^0)$ determines the combination
\beq
\left[2 \nabla^i\chi_i - \p_\rho(\lambda^i_i)\right]=F\left(\rho,v_i,P,\phi_i\right)
\eeq
where the known but unilluminating function on the right hand side (given for the Rindler case in \cite{bkls}) is regular in $\rho$. Provided the above conditions are satisfied, the perturbed geometry satisfies all the Einstein equations through $\mathcal{O}(\y^0)$ and the prescribed boundary conditions on \sc\ (conformal flatness and constant $K$). If we want to solve the Einstein equations at higher order it may be necessary to add corrections to specific metric components at $\mathcal{O}(\y)$ or higher which do not affect the Einstein equations through $\mathcal{O}(\y^0)$.

Since the left hand side of (\ref{phie}) is a total divergence a potential obstruction to solving the equation may arise by integrating over the  $p$-sphere in \sc . This integral gives
\beq \label{zm}
\p_t\left(\int v^2 d\Omega_p\right)=-p\p_t\left(\int P d\Omega_p\right).
\eeq
The incompressible Navier-Stokes equation has a shift symmetry under which 
\be v_i(x,t)\to v_i(x,t),~~~P(x,t)\to P(x,t)+f(t),\ee
where $f$ is any function of time.  That is, the spatially independent zero mode of the pressure does not enter the Navier-Stokes equation. We can exploit this shift freedom to solve (\ref{zm}).

Interestingly, the right hand side of (\ref{zm}) is absent for the pure Dirichlet boundary conditions originally employed in \cite{bkls}.  Hence
there is a higher-order obstruction to solving the equations with these boundary conditions: one cannot keep the total area of \sc\ fixed. Perhaps this is somehow related via the Bekenstein-Hawking entropy-area law to entropy production in the fluid. Indeed the left hand side of (\ref{zm}) is proportional to the total energy dissipation of the fluid. 

In reference \cite{Compere:2011dx} it was proven that the dual-fluid solutions with Rindler horizons constructed to third order in \cite{bkls} exist to all orders in the near-horizon expansion parameter $\y$. We have not given a proof of the analogous result for the deformed Schwarzschild horizons considered here. Potential obstructions exist from integrating over the $p$-sphere, but at least the most obvious of these can be made to vanish by adjusting the time dependent zero mode of the pressure at every order in $\y$.  A generalization of the proof of \cite{Compere:2011dx} to the Schwarzschild case would be of great interest.

\section{Cosmic censorship and global existence}

  The cosmic censorship conjecture of general relativity posits that, under suitable conditions,  singularities on or outside event horizons do not arise from smooth initial data.  The global existence conjecture for the incompressible Navier-Stokes equation similarly posits that singularities never form from smooth initial velocity fields. 
  
  Having found a precise mathematical relation between the Einstein and the Navier-Stokes equations, in which the Navier-Stokes fluid lives on a surface outside the horizon, it seems within the realm of possibility to find a precise relation between some form of the two conjectures\footnote{ This is more natural in the present setup in which the Einstein solutions under consideration are typically asymptotically flat in all spatial directions, as opposed to the planar situations considered in \cite{bkls}.} \cite{Oz:2010wz}. We shall not attempt to find such a relation in the present work but will make a few simple observations. 
  
  Suppose we had a counterexample to global existence of the Navier-Stokes equation on the $p$-sphere in which one begins with a smooth velocity field and then develops a singularity in some finite time. Our construction then gives a perturbative solution of the $p+2$ dimensional Einstein equation that looks like a black hole with some disturbances outside the horizon. Let us moreover assume that the perturbation expansion has a finite radius of convergence in $\alpha$ (this of course could be difficult to actually prove). We then have a solution of the Einstein equation which is evolved from smooth initial data and has a singularity. The singularity is outside the horizon because \sc\ is outside the horizon. So a counterexample to global existence might thus be used to construct a counterexample to some form of cosmic censorship. It is not the usual form of cosmic censorship considered for example in asymptotically flat spacetimes however because we have imposed Dirichlet-like boundary conditions on \sc. 
Nevertheless this example illustrates the possibility of a potentially fruitful connection between the two conjectures. 

\section{Acknowledgements}
We are grateful to Lars Andersson, Sean Hartnoll, Cindy Keeler, Slava Lysov, Gim Seng Ng and Brian Shuve for helpful conversations.

  \end{document}